\documentclass[conference]{IEEEtran}
\makeatletter
\def\ps@headings{%
    \def\@oddhead{\mbox{}\scriptsize\rightmark \hfil \thepage}%
    \def\@evenhead{\scriptsize\thepage \hfil \leftmark\mbox{}}%
    \def\@oddfoot{}%
    \def\@evenfoot{}}
\makeatother
\usepackage[top=0.76in, bottom=1in, left=0.63in, right=0.66in]{geometry}
\usepackage{paralist}
\usepackage{pgf}
\usepackage{epstopdf}
\usepackage{etex}
\usepackage{tikz}
\usetikzlibrary{arrows,automata}
\usepackage{algcompatible}
\usepackage{algorithm}
\usepackage{soul}
\usepackage{amsmath}

\usepackage{amssymb}
\usepackage{tikz-qtree}
\usepackage{array}
\usepackage[american]{babel}
\usepackage{graphicx}
\usepackage{subfigure}
\usepackage{amssymb}
\usepackage{enumitem}
\usepackage{algorithm}
\usepackage{url}
\usepackage{xcolor,listings}
\usepackage{pgfplots}
\usepackage{verbatimbox}
\usepackage{multirow}
\usepackage[section]{placeins}
\usepackage{tikz}

\pgfplotsset{every axis legend/.style={
        cells={anchor=center},% Centered entries
        inner xsep=3pt,
        inner ysep=2pt,
        nodes={inner sep=2pt,text depth=0.15em},
        anchor=south east,
        shape=rectangle, %box shape
        fill=white, %box inside
        draw=white, %box color
        at={(rel axis cs:0.50,0.65)} %box position right top
}}

\usepackage{amsthm}
\usepackage{algorithm}
\usepackage[noend]{algpseudocode}
\usepackage{forest}

\usetikzlibrary{matrix,calc,positioning}
\usetikzlibrary{arrows.meta, shapes.geometric, calc, shadows}

\colorlet{mygreen}{green!75!black}
\colorlet{col1in}{red!30}
\colorlet{col1out}{red!40}
\colorlet{col2in}{mygreen!40}
\colorlet{col2out}{mygreen!50}
\colorlet{col3in}{blue!30}
\colorlet{col3out}{blue!40}
\colorlet{col4in}{mygreen!20}
\colorlet{col4out}{mygreen!30}
\colorlet{col5in}{blue!10}
\colorlet{col5out}{blue!20}
\colorlet{col6in}{blue!20}
\colorlet{col6out}{blue!30}
\colorlet{col7out}{orange}
\colorlet{col7in}{orange!50}
\colorlet{col8out}{orange!40}
\colorlet{col8in}{orange!20}
\colorlet{linecol}{blue!60}

\usepackage{listings}
\usepackage{xcolor}
\usepackage{float}
\usepgfplotslibrary{groupplots}
\colorlet{punct}{red!60!black}
\definecolor{background}{HTML}{EEEEEE}
\definecolor{delim}{RGB}{20,105,176}
\colorlet{numb}{magenta!60!black}

\lstdefinelanguage{json}{
    basicstyle=\normalfont\ttfamily,
    numbers=left,
    numberstyle=\scriptsize,
    stepnumber=1,
    numbersep=8pt,
    showstringspaces=false,
    breaklines=true,
    frame=lines,
    backgroundcolor=\color{background},
    literate=
     *{0}{{{\color{numb}0}}}{1}
      {1}{{{\color{numb}1}}}{1}
      {2}{{{\color{numb}2}}}{1}
      {3}{{{\color{numb}3}}}{1}
      {4}{{{\color{numb}4}}}{1}
      {5}{{{\color{numb}5}}}{1}
      {6}{{{\color{numb}6}}}{1}
      {7}{{{\color{numb}7}}}{1}
      {8}{{{\color{numb}8}}}{1}
      {9}{{{\color{numb}9}}}{1}
      {:}{{{\color{punct}{:}}}}{1}
      {,}{{{\color{punct}{,}}}}{1}
      {\{}{{{\color{delim}{\{}}}}{1}
      {\}}{{{\color{delim}{\}}}}}{1}
      {[}{{{\color{delim}{[}}}}{1}
      {]}{{{\color{delim}{]}}}}{1},
}

\ifCLASSINFOpdf
\else
\fi

\begin{document}
\newtheorem{mydef}{Definition}

\title{SDN based Network Function Parallelism in Cloud}

\author{
    \IEEEauthorblockN{Ankur Chowdhary, Dijiang Huang}
    \IEEEauthorblockA{Arizona State University
        \\\{achaud16, dijiang\}@asu.edu}
}
%\and
%\IEEEauthorblockN{James Kirk\\ and Montgomery Scott}
%\IEEEauthorblockA{Starfleet Academy\\
%San Francisco, California 96678--2391\\


%Telephone: (800) 555--1212, Fax: (888) 555--1212}


\maketitle

\begin{abstract}

%Security provisioning for a large cloud network is a challenging task. Software-defined network (SDN) has emerged as a framework for security policy enforcement in a cloud network. We propose an SDN based distributed security framework for cloud network. We use flow aggregation and security function dependency parallelization to create service function chaining framework that scales well on the large cloud network. We have utilized the datapath development kit (DPDK) with OpenFlow switch to enhance packet processing for security functions. Our framework shows a 1.41-1.92x reduction in mean latency for service function chaining of security operations. 

%We present a scalable security framework for SDN enabled cloud networks based on Service Function Chaining (SFC) and parallelized packet processing. Our framework identifies operational dependencies between security service functions (SF's) and parallelizes operations using datapath development kit (DPDK). We use queuing model $M/M/c$ queue to model SF's and analyze our optimal chaining of security functions in a multi-tenant SDN cloud network.  

%We have used a queuing model to analyze the effectiveness of our optimal SFC architecture.

%Our architecture considers various network and application security functionalities such as firewall, network address translation (NAT), intrusion detection system (IDS) as service functions (SF) and optimizes service function path (SFP) while applying various SF’s between two nodes in a network. 
Network function virtualization (NFV) based service function chaining (SFC) allows the provisioning of various security and traffic engineering applications in a cloud network. Inefficient deployment of network functions can lead to security violations and performance overhead. In an OpenFlow enabled cloud, the key problem with current mechanisms is that several packet field match and flow rule action sets associated with the network functions are non-overlapping and can be parallelized for performance enhancement. We introduce Network Function Parallelism (NFP) \texttt{SFC-NFP} for OpenFlow network. Our solution utilizes network function parallelism over the OpenFlow rules to improve SFC performance in the cloud network. We have utilized the DPDK platform with an OpenFlow switch (OVS) for experimental analysis. Our solution achieves a 1.40-1.90x reduction in latency for SFC in an OpenStack cloud network managed by the SDN framework.

%Existing SFC approaches identify component dependencies and utilize parallelism for performance improvement in NFV. In an OpenFlow enabled cloud, the key problem with current mechanisms is that several packet field match and flow rule action sets associated with the network functions are overlapping. 

\end{abstract}

\IEEEpeerreviewmaketitle

\section{Introduction}
\noindent The real-time ad-hoc provisioning of SFC in a multi-tenant cloud network is a challenging task. Software Defined Network (SDN) allows orchestration and centralized management of a multi-tenant cloud network. SDN separates data plane and control plane in a network, so that network switches become simple forwarding devices. There are several security features such as firewall, network address translation (NAT), deep packet inspection (DPI), intrusion detection and prevention system (IDPS) which can be deployed efficiently in an SDN based cloud network. 

\noindent These security and network functions can be virtualized using network function virtualization (NFV) approach. Instead of a dedicated hardware resources such as routers, firewalls, and switches, the NFV~\cite{li2015software} allow software-based implementation of various security network functions called virtual network functions (VNF's).  \textit{Service Function chaining (SFC)} provides an ordered list of VNF's to act in serial or parallel order providing optimal security in SDN enabled cloud network.

\noindent There are research works that consider the implementation cost, throughput~\cite{xu2017approximation} and mean response time~\cite{prados2017analytical} for deployment of VNF's for Optimal SFC. A VNF intrusion detection system (IDS) before VNF firewall is sub-optimal since the firewall can filter out part of traffic, which needs to be mirrored for intrusion detection thus reduce traffic volume to be mirrored for intrusion detection. Additionally, some network functions such as IDS and network-probe involve only mirroring or passive monitoring of network traffic. Since they involve no modification to the packet header, they can be operated in parallel. This reduces the performance hit induced by serial processing of VNF's. Current works, however, fail to identify opportunities of parallelization of VNF which have a non-overlapping match and action sets in SFC environment. 

\noindent In this research work, we compile the network functions into OpenFlow rules in order to check the overlap between a match and action fields of network traffic. \texttt{SFC-NFP} achieves gain in performance by up to 48\%~\cite{luo2015practical}. VNF's operate in parallel using data path development kit (DPDK)~\cite{dpdk2} based Openflow~\cite{openflow} switches. Our experimental results show packet processing performance enhancement by 2.5x times~\cite{dpdk1}. We make use of flow aggregation and network function parallelism for performance optimization and security enforcement in the SDN based cloud network.

The key contribution of this research work are:-
\begin{enumerate}
\item Identification of overlap between match and action fields of various service functions in cloud network.
\item Performance optimization by in SFC using network function parallelism. This will ensure reduced SFC latency and higher throughput which is critical for security in a cloud network. 
 
\end{enumerate} 

%The remainder of the paper is organized as follows. In section II, we discuss background details about security and performance issues in SFC for a cloud network, DPDK framework to enhance packet processing for SDN environment and use of M/M/c queue model for SFC.  We also discuss details about parallelization of network functions. The details of the system model and architecture have been discussed in section III. Optimal network function chaining using M/M/c queue model has been discussed in section IV. We also introduce parallelization of network functions in this section. The details of the experimental setup and results of an end to end latency for SFC has been discussed in section V. We discuss the related works and compare our approach to work done in the field of service function chaining, performance and security optimization approaches in SDN and NFV in section VI. Finally, we conclude our work in section VII and provide details of future work for secured SFC in SDN environment.
  
%http://dpdk.org/doc/guides/prog_guide/poll_mode_drv.html#hardware-offload

\section{Background}

\begin{figure}[!ht]
    \centering
    \includegraphics[width=0.5\textwidth]{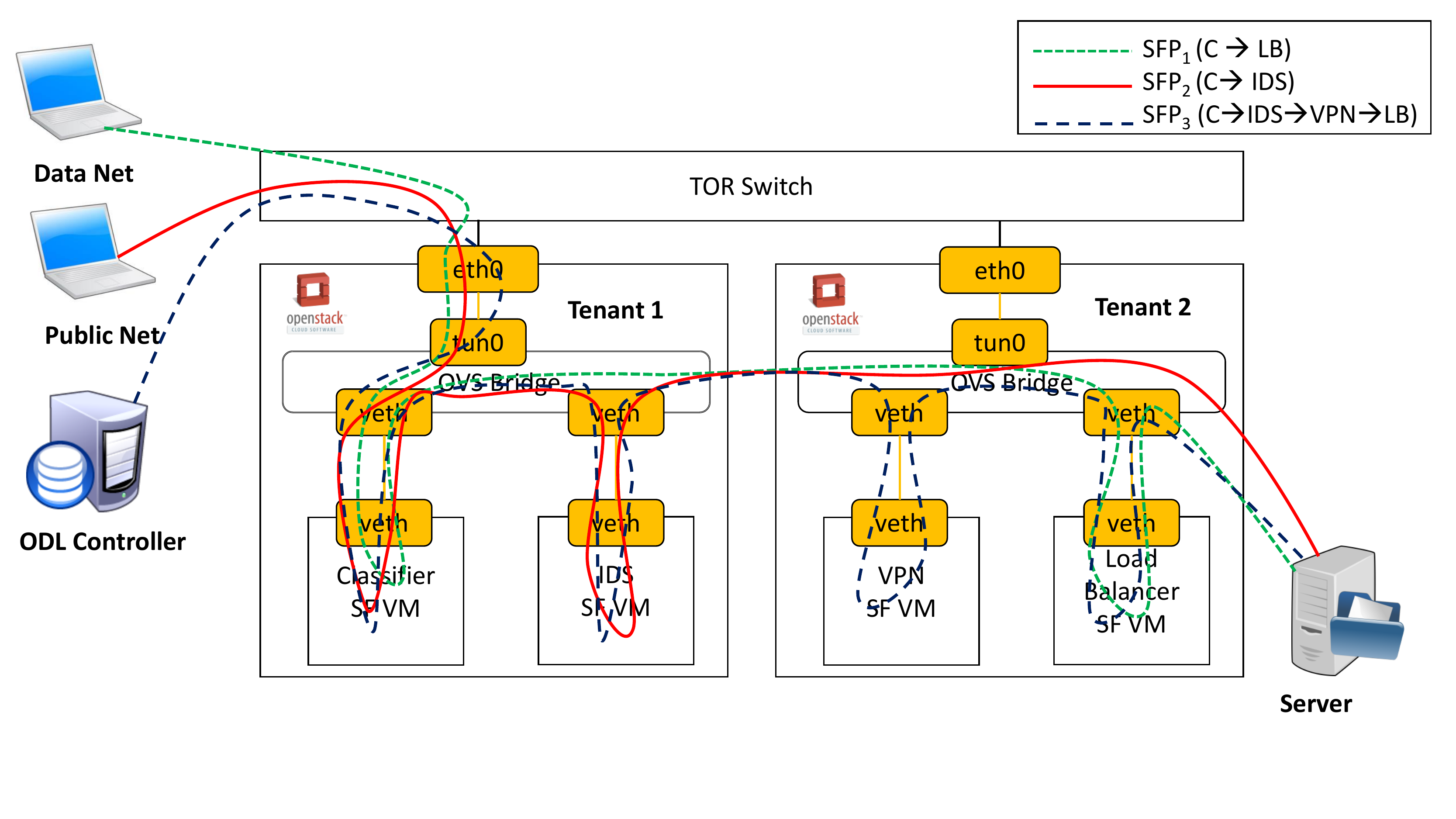}
    \caption{SFC in Multi-Tenant Cloud Network}
     \label{fig:multi-tenant}
\end{figure}

\noindent In a multi-tenant cloud network, different tenants can have different security requirements, e.g., $Tenant_1$ in Figure~\ref{fig:multi-tenant} can have a security requirement of firewall at the gateway of the network in order to ensure users outside the cloud network cannot directly communicate with cloud virtual machines ($VM_1, VM_2$). Similarly, communication across two different tenants ($Tenant_1, Tenant_2$) as shown in Figure~\ref{fig:multi-tenant} distributed geographically may require a company mandated virtual private network (VPN).

\noindent We consider all such security features as VNF's. The SFC in effect creates an overlay between network nodes (VM's) independent of network topology. This scheme allows connection of a large number of VNF's in NFV environment. These connections are ad-hoc and can be setup or torn on a need basis. A \texttt{Service Function Forwarder (SFF)} is a layer forwarding provider to deliver packets from SFC to destination nodes in the network. 

\noindent The OVS for each tenant acts as SFF in Figure~\ref{fig:multi-tenant}. \texttt{Service Function Path (SFP)} is the VNF's traversed by network traffic for packet delivery from source to destination node in cloud network. In Figure~\ref{fig:multi-tenant} SFP for communication between external users (Ext-NET) and virtual machines of $Tenant_1$ is Ext-NET $\rightarrow IDS \rightarrow Firewall \rightarrow VM_1$.   

\begin{table}[!ht]
    \begin{center}
        \begin{tabular}{ | m{1.1cm} | m{7cm}| }
            \hline
            \textbf{VNF} & \textbf{Security Policy} \\
            \hline
             IDS         & \texttt{1) Ext-NET any $\rightarrow$ 10.1.0.0/24 any alert("DDoS")} \\
                         & \texttt{2) 10.1.0.0/24 80 $\rightarrow$ 192.168.1.0/24 80 alert("Buffer Overflow")} \\
            \hline
            Firewall     & \texttt{1) iptables -A INPUT -i eth1 -p tcp -s Ext-NET --dport 80 -j DROP} \\
            \hline
            VPN          & \texttt{1) $Tenant_1$ -i eth1 $\rightarrow$ $Tenant_2$ -i eth0 - OpenVPN } \\
            \hline
        \end{tabular}
    \end{center}
    \caption{ \label{tab:tab00} VNF Security Policies in Cloud Network}
      \vspace{-2.0em}
\end{table}

\noindent Consider the VNF IDS for $Tenant_1$ - Rule 1 in Table~\ref{tab:tab00}, where any traffic from Ext-NET coming to an internal network (10.1.0.0/24) needs to be examined for distributed denial of service (DDoS) attack pattern. Also the VNF firewall rule on $Tenant_1$ blocks any traffic from external users on port 80 i.e., web service hosted on $Tenant_1$ VM's can only be accessed by internal users. The IDS will, however, inspect web traffic as well for DDoS attack pattern. Since firewall rule already blocks that traffic, it is redundant to examine web traffic by IDS, if the SFP is instead \texttt{Ext-NET $\rightarrow Firewall \rightarrow IDS \rightarrow 10.1.0.0/24$}, this is much more efficient from a performance perspective.

\noindent On the other hand VNF IDS - Rule 2 which is applicable for traffic between $Tenant_1$ and $Tenant_2$. The SF followed by SFC is \texttt{$10.1.0.0/24 \rightarrow VPN \rightarrow IDS \rightarrow 192.168.1.0/24 $}. The IDS rule 2 in Table~\ref{tab:tab00} should ideally examine traffic from $Tenant_1$ to $Tenant_2$ for buffer overflow attack patterns. However, since VNF VPN encrypts traffic between both tenants, the IDS will not be able to check the payload and packet headers of the traffic. Thus, traffic needs to be mirrored to IDS before being encrypted by VPN VNF. The desired SFP for security provisioning is \texttt{$10.1.0.0/24 \rightarrow IDS \rightarrow VPN \rightarrow 192.168.1.0/24 $}. 

\noindent We utilize the flow rule compilation of VNF ruleset so that VNF's can be expressed in the common OpenFlow rule format. The OpenFlow rules are checked for match and action set overlap in our framework. Once the flow rules are aggregated, we use DPDK based parallelized packet processing
for fast packet processing and security provisioning.

\noindent SDN architecture relies on OpenFlow protocol~\cite{openflow} for communicating with OpenFlow enabled network switches. An OpenFlow switch consists of one or more flow tables for packet lookup and forwarding. Each table of the OpenFlow switch consists of entries - match fields, counters, and instruction set that apply to the matching packet. The packet processing in the current Open vSwitch (most commonly used OpenFlow switch) architecture occurs using daemon \textit{ovs-vswitchd} which lies in kernel space. This poses two main issues, both of which slows down throughput and limit the scalability of the architecture. 
\begin{enumerate}
\item System calls are required for context switch when flow matching and action operations take place in OpenFlow network. 
\item Packet processing occurs in serial i.e., dedicated queue for every network interface card (NIC).
\end{enumerate}

\noindent The DPDK drivers make use of Non-unified memory architecture (NUMA), which avoids dedication of a separate CPU     core for transmission (TX) and reception (RX) of network packets on a NIC~\cite{dpdk1}. Each NIC can have multiple TX/RX queues which allow us to perform SFC for various security VNF's in parallel. 
 
%\subsection{DPDK based Fast Packet Processing}

%The DPDK framework is comprised of libraries for complex packet processing applications. 
%\begin{figure}[!ht]
%  \centering
%    \includegraphics[width=0.5\textwidth]{dpdk03.pdf}
%     \caption{DPDK based fast packet processing in SDN}
%     \label{fig:dpdk4}
%\end{figure}
%As shown in the Figure~\ref{fig:dpdk4}, the DPDK makes use of poll mode driver (PMD)~\cite{pmd}. PMD accesses RX and TX descriptors directly without any interrupts so that packets can be %processed and delivered quickly to user applications. Compared to native OVS architecture where packet processing operations require a context switch to kernel space, the DPDK driver processes packets in userspace entirely hence accelerates performance. We have utilized DPDK with OVS in our system to optimize packet processing and utilize the multi TX and RX queues for parallelizing network functions independently of each other. 

%https://software.intel.com/en-us/articles/introduction-to-the-data-plane-development-kit-dpdk-packet-framework
%http://dpdk.org/doc/guides/prog_guide/poll_mode_drv.html
\subsection{SFC Parallelization Model}

\noindent For each security operation like IDS, firewall, etc. we utilize one $M/M/c$ queue for a theoretical estimate of parallel packet processing. The parameters M represents the packet arrival rate $\lambda$ and packet service rate $\mu$. The parameter 'c' represents the number of servers handling the packets. Since each TX/RX buffer allows parallel packet processing, we can have upto 'c' serving threads for each security operation. The packet arrival rate $\lambda$ is modeled using Poisson's distribution in $M/M/c$ model and the packet service rate $\mu$ is modeled using exponential distribution. 

\noindent We explore the parallelization between network functions by checking the order of dependency between functions. For $\textit{DROP}$ action associated with one network function firewall can affect the functionality of the subsequent network function $\textit{Load Balancer}$. Since the packet processing of most security functions depends on either read/write operations on header or packet payload, we identify these dependencies while constructing parallel packet processing queues. There can be following relations between network functions - Read after Read (RAR), Write after Read (WAR), Read after Write (RAW) and Write after Write (WAW). The network functions $SF_1$ and $SF_2$ can be parallelized if dependencies between them are Read after Read (RAR) or Write after Read (WAR). The table below shows operations on packet header and payload by various VNF's.  
%We consider the different layers of the IP header as network functions. Figure~\ref{fig:mmc01} below shows that on receiving $packet\_in$ event the OpenFlow switch parses various network functions, e.g. $SF_{L2}$ depicts the source and destination MAC addresses, $SF_{L4}$ depicts source and destination TCP/UDP port.

%\begin{figure}[!ht]
%  \centering
%    \includegraphics[width=0.45\textwidth]{diag02.pdf}
%     \caption{Service Function Chaining based on  M/M/c Queue Model}
%     \label{fig:mmc01}
%\end{figure}

%The SFC for IDS comprises of SF's from layers 2,3,4 $SFC_{IDS} = \{SF_{L2}, SF_{L3}, SF_{L4}\}$. Similarly $SFC_{Firewall} = \{SF_{L3}, SF_{L4}\}$. The SFC output via TX buffer of one Open vSwitch (OVS) can act as input for $SFC_{DPI}$ via RX buffer of second OVS in the Figure~\ref{fig:mmc01}. Thus $SFC_{DPI}$ on the second Openflow switch doesn't need to recompute the $SFC_{IDS}$ for layers 2-4. It will only need $SF_{L5-L7}$ to be appended to $SFC_{IDS}$. This will allow faster packet processing for network function path from $packet\_in$ to $packet\_out$. Additionally, each Openflow switch utilizes DPDK netdev driver, which transmits packets from userspace to NIC's of each OVS. This allows a significant throughput increase. 

%\subsection{Parallelization of Network Functions}

\begin{table}[!ht]
\begin{center}
\begin{tabular}{ | m{4em} | m{3em} | m{4em}| }
\hline
\textbf{Function } & \textbf{Header} & \textbf{Payload}\\ 
\hline
Probe & R & - \\ 
\hline
NAT   & R/W & - \\
\hline
Firewall & R/W & - \\
\hline
Proxy & R & R \\
\hline
IDS & R & R \\
\hline
IPS & R/W & R \\
\hline
Load Balancer & R/W & R \\
%\hline
%VPN & R/W & R \\
\hline
\end{tabular}
\end{center}
\caption{\label{tab:tab01} Network functions and packet operations}
  \vspace{-1.5em}
\end{table}

As can be observed from Table~\ref{tab:tab01}, probe, and NAT operations are RAR for packet header. Also, IDS and Firewall operations are WAR, so these VNF's can be parallelized.  Whereas the operations that involve WAW on packet header or payload such as load balancers and IPS cannot be parallelized. We implement service function chaining (SFC) algorithm at application plane using ODL controller. The virtual network functions such as group-based policy (GBP), Firewall, deep packet inspection (DPI), etc. are stitched as a part of the service function queue using optimized service function chaining algorithms discussed in next sections.

\section{Network Function Parallelism M/M/c Queue Model}
\noindent We use a queuing model for the M/M/c queue for optimizing the service function allocation for every service function chain (SFC). We define key parameters for an M/M/c queue in the table below. 

\begin{table}[!ht]

\begin{center}
\begin{tabular}{ | m{3.5em} | m{5.5cm}| } 
\hline
\textbf{Variable} & \textbf{Description}\\ 
\hline
n & Number of service functions \\ 
\hline
$p_n$ & Equilibrium Probability for n VNF's\\
\hline
$\lambda$ & Packet arrival rate \\
\hline
$\mu$ & Packet service rate \\
\hline
$\rho$    & $\frac{\lambda}{\mu}$ \\
\hline
c     & Number of servers \\
\hline 
$p_0$ & Equilibrium Probability for zero VNF's\\
\hline
$E(W)$ & Mean wait time for VNF in queue \\
\hline
\end{tabular}
\end{center}
\caption{\label{tab:tab02} M/M/c queue parameters}
  \vspace{-2em}
\end{table}

\begin{figure}[!ht]
  \centering
    \includegraphics[trim={0cm 3cm 0cm 3cm}, clip, width=0.5\textwidth]{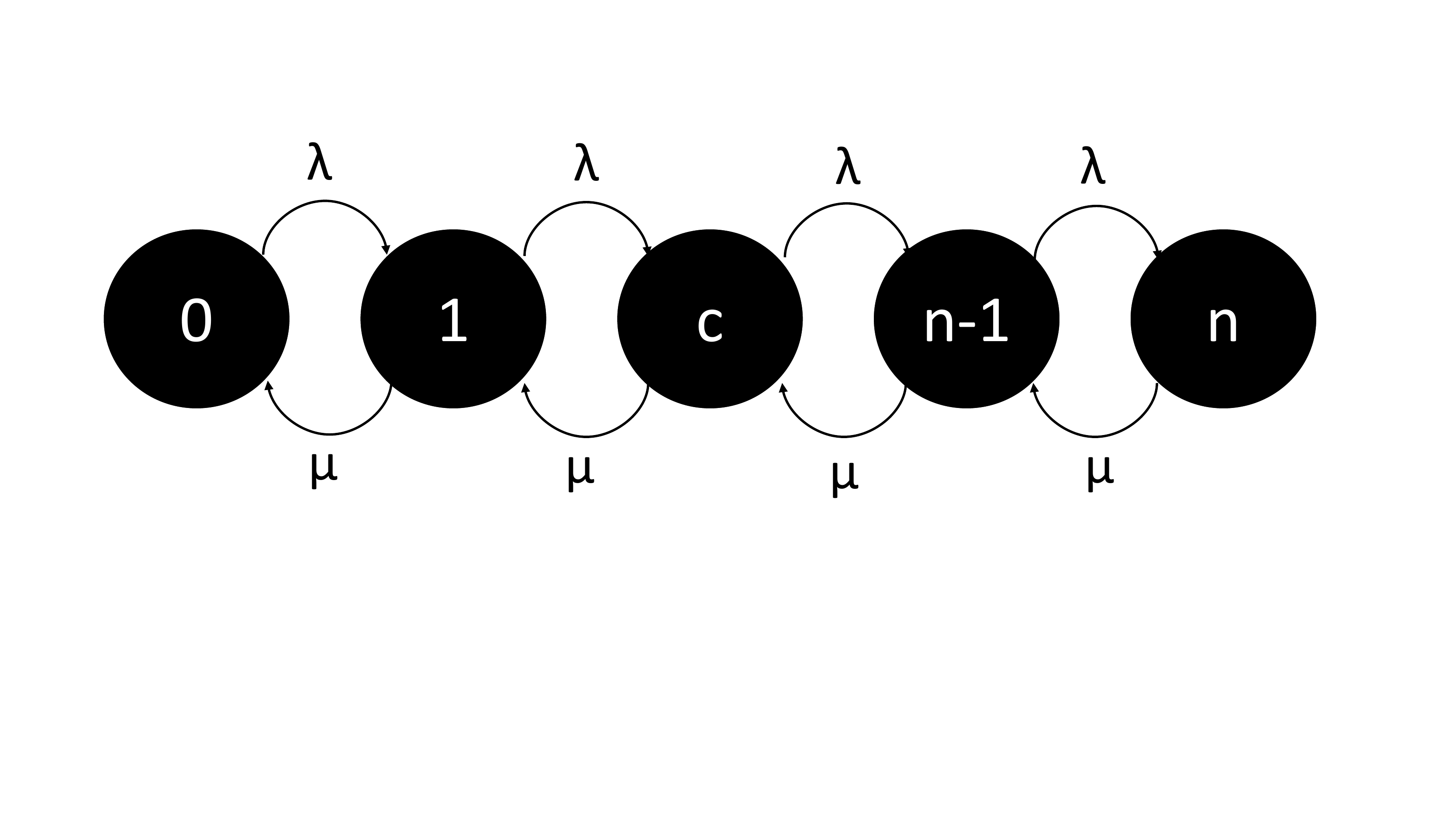}
    \vspace{-4em}
     \caption{Flow Diagram for M/M/c queue}
     \label{fig:mmc02}
\end{figure}

Consider flow diagram Figure~\ref{fig:mmc02}, where 'n' network functions are arriving to be processed by SFC. 

The flow exchange between two neighboring states is given by 

\begin{equation}
\lambda \times p_{n-1} = min. (n,c) \mu \times p_n 
\end{equation}
\begin{equation}
p_n = \frac{(cp)^n}{n!} \times p_0
\end{equation}
\begin{equation}
p_{c+n} =     \rho^n \times p_c = \rho^n\frac{(c\rho)}{c!}\times p_0    
\end{equation}

The normalization for equations above give
\begin{equation}
p_0 = \bigg(  \sum_{n=0}^{c-1} \times \frac{(c\rho)^n}{n!} + \frac{(cp)^c}{c!} \times \frac{1}{1-\rho} \bigg)^{-1} 
\end{equation}    
The probability that a given VNF has to wait to be executed in a SFC is denoted by delay probability $\Pi_W = p_c + p_{c+1} + p_{c+2}..$. 

\begin{equation}
\Pi_W = \frac{(cp)^c}{c!} \bigg( (1-\rho) \sum_{n=0}^{c-1} \times \frac{(c\rho)^n}{n!} +  \frac{(cp)^c}{c!}  \bigg)^{-1} 
\end{equation} 

The mean length of queue is calculated using equilibrium probability
\begin{equation}
\begin{split}
E(L^q) = \sum_{n=0}^{\infty} np_{c+n} = \frac{p_c}{1-\rho} \sum_{n=0}^{\infty} n(1-\rho)\rho^n \\
\end{split}
\end{equation}

\begin{equation}
E(L^q) = \Pi_W.\frac{\rho}{1-\rho}   
\end{equation}

\noindent The queue length will give an estimation of network functions waiting to be processed thus new VNF can be chained based on mean waiting time. We can make use of Little's Law to obtain mean waiting time for a virtual network function in SFC.

\begin{equation}
E(W) = \Pi_W.\frac{1}{1-\rho}.\frac{1}{c\mu}
\end{equation}

The time between two successive network functions being processed is a minimum of 'c' exponential service times with mean $\frac{1}{\mu}$. This value can also be represented as exponential time with mean $\frac{1}{c\mu}$. Thus,

\begin{equation}
E(W) = \Pi_W.\frac{1}{c\mu} + E(L^q) \frac{1}{c\mu}
\end{equation}
\noindent The real-time security provisioning will need minimal waiting time for critical services such as Intrusion Prevention System (IPS), as opposed to Quality of service (QoS). The mean waiting time estimated using queuing model has been used in the experimental analysis for comparing flow optimization performance gain against mean waiting time when VNF's are parallelized using dependencies of read/write between them discussed in Section II-A. We discuss prioritized network function provisioning algorithm in the next subsection. This algorithm optimizes the processing of functions in an SFC. 

\subsection{Network Function Parallelization}
 \noindent The algorithm \textit{Network Function Parallelization (NFP)} checks the action list of each flow rule line \textit{4}. If the action associated with flow rule involves forwarding (fwd) for the matching traffic, \textit{$(fwd, fwd)$} or  \textit{$(fwd, flow\_mod)$} line \textit{6}.
 
\begin{algorithm}
    \caption{Network Function Parallelization}\label{euclid}
    \begin{algorithmic}[1]
        \Procedure{Network Function Parallelization}{\texttt{F}}
        \State \texttt{$F \gets$ current flow rules}
        \For{\texttt{$i \in$ \{1,n\} }}
        \State \texttt{actions $\gets$ $f_i$.actions()}
        %\For{\texttt{a $\in$ actions}}
        \For {\texttt{$j,k \in$ \{1,n\} }} 
        \If {\texttt{$a_j,a_k$ $\in$ (fwd,fwd) or (fwd, flow\_mod)}}
        \State Parallelize ($a_j,a_k$)             
        \ElsIf  {\texttt{$a_j,a_k$ $\in$ (flow\_mod,fwd) or (flow\_mod, flow\_mod)}}
         \State Serial ($a_j,a_k$)    
        \EndIf
        \EndFor
        \EndFor    
        \EndProcedure
    \end{algorithmic}
\end{algorithm} 
 
\noindent The parallelization between network functions related to action , e.g. conntrack and NAT is possible, however, if flow\_mod operation occurs before forward in two actions for a traffic match or \textit{$(flow\_mod, flow\_mod)$} for two actions occurs for a particular flow rule - line \textit{8}, the network functions are required to be operated in serial order, e.g. NAT and Intrusion Prevention System (IPS) for a flow match will occur in serial order.

\section{Implementation and Evaluation}
\noindent We utilized an OpenStack based cloud network comprising of two Dell R620 servers and two Dell R710 servers all hosted in the ASU data center. Each
Dell server has about 128 GB of RAM and 16 core CPU. The SDN controller Opendaylight-Carbon has been used for network management and orchestration. 
The Openstack version Ocata was utilized for implementation of DPDK enabled architecture. The script $\textit{networking-ovs-dpdk}$ was incorporated in neutron component of Openstack to enable support for DPDK. The OVS version 2.8.90 was used on compute nodes for Openstack cloud and datapath was modified to use $\textit{netdev}$ driver. 
\begin{table}[!ht]
\begin{center}
\begin{tabular}{ | m{3.0cm} | m{3.0cm}| }
\hline
\textbf{Security Function} & \textbf{Software} \\ 
\hline
Probe & nmap \\ 
\hline
NAT   & netfilter \\
\hline
Firewall & netfilter \\
\hline
IDS & Snort \\
\hline
Load Balancer and Proxy & nginx \\
\hline
VPN & openvpn \\
\hline 
DPI & OpenDPI \\
\hline
\end{tabular}
\end{center}
\caption{\label{tab:tab03} Service functions and Softwares}
  \vspace{-2em}
\end{table}
\noindent We utilized the following security, virtual network functions and corresponding software shown in Table~\ref{tab:tab03} for experimental evaluation. We created a multi-tenant network in an OpenStack cloud environment for experimental evaluation. The Ubuntu Mini OS was used for guest VM's that were connected to the OVS in each compute node allow the guest VM to become a part of SDN network.

\subsection{SFC Latency}
\noindent We measured the latency for the qperf tool to evaluate the end to end latency when NAT, Firewall, and IDS are used as security VNF's on top of OVS. The experiment evaluates the effect of using NFP \texttt{SFC-NFP} for providing VNF's as opposed to VNF's in serial \texttt{SFC-Serial} which assumes that support of OVS-DPDK is missing in the deployment of VNF's for SDN cloud network. The formula for theoretical estimates of multi-core architecture latency \texttt{SFC-Theoretical} using M/M/c queuing model has been discussed in Section III.   

%\begin{figure}[!ht]
%    \centering
%    \includegraphics[width=0.45\textwidth]{nfp01.pdf}
%    \caption{SFC latency analysis with Firewall, NAT and IDS}
%     \label{fig:expt1}
%\end{figure}
%As show in the Figure~\ref{fig:expt1}, we assess the effect of using VNF's in serial and reduction in latency when parallelization is introduced for fast packet processing \texttt{SFC-Theorectical}. Our method uses flow rule compilation and aggregation in addition to parallel packet processing in SDN enabled cloud \texttt{SFC-NFP}.

\begin{figure*}
  \centering
\begin{tikzpicture}[scale=0.9]

\begin{groupplot}[group style={group size=3 by 1}, height=6cm,width=6.3cm, legend style={font=\tiny},xtick={50,100,150,200,250}, ytick={2,4,6,8,10,12} ] 

\nextgroupplot[xlabel={network-size}, ylabel={Latency (s)}, ylabel style={yshift=-0.7cm}, title=a) OVS-DPDK 2 Cores]

\addplot coordinates { (250, 10.1) (200, 4.87) (150, 1.58) (100, 0.98) (50, 0.45)}; 
 \addlegendentry{SFC-Serial}

\addplot coordinates { (250, 9) (200, 4.26) (150, 1.53) (100, 0.67) (50, 0.28)}; 
\addlegendentry{SFC-Theoretical}

\addplot[color=green,
mark=triangle,] coordinates { (250, 6.05) (200, 2.72) (150, 1.12) (100, 0.56) (50, 0.23)}; 
\addlegendentry{SFC-NFP}

 \nextgroupplot[xlabel={network-size}, ylabel={Latency (s)}, ylabel style={yshift=-0.7cm},  title=b) OVS-DPDK 4 Cores]
 
\addplot coordinates { (250, 10.1) (200, 4.87) (150, 1.58) (100, 0.98) (50, 0.45)}; 
\addlegendentry{SFC-Serial}

\addplot coordinates { (250, 6.1) (200, 2.96) (150, 1.03) (100, 0.57) (50, 0.18)}; 
\addlegendentry{SFC-Theoretical}

\addplot[color=green,
mark=triangle,] coordinates { (250, 4.05) (200, 1.62) (150, 0.72) (100, 0.41) (50, 0.13)}; 
\addlegendentry{SFC-NFP}

\nextgroupplot[xlabel={network-size}, ylabel={Latency (s)}, ylabel style={yshift=-0.7cm}, title=c) OVS-DPDK 8 Cores ]

\addplot coordinates { (250, 10.1) (200, 4.87) (150, 1.58) (100, 0.98) (50, 0.45)}; 
\addlegendentry{SFC-Serial}

\addplot coordinates { (250, 4.1) (200, 2.07) (150, 0.58) (100, 0.28) (50, 0.12)}; 

\addlegendentry{SFC-Theoretical}

\addplot[color=green,
mark=triangle,] coordinates { (250, 2.95) (200, 1.82) (150, 0.46) (100, 0.21) (50, 0.10)}; 
\addlegendentry{SFC-NFP}

%\addlegendentry{SFC-FO}

\end{groupplot}

\end{tikzpicture}
 \caption{SFC Latency}
      \label{fig:expt4}
\end{figure*}
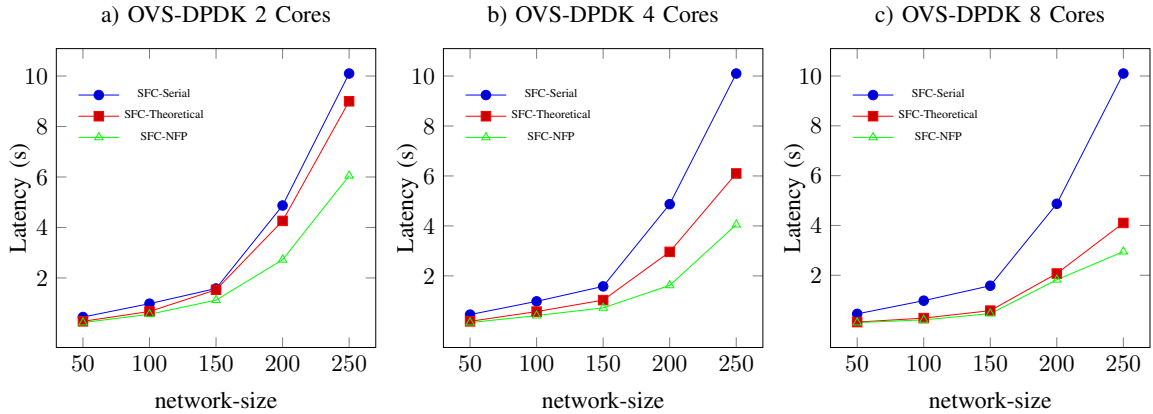

\noindent We increase the number of CPU core from 2 to 8 for the experiment as shown in Figure~\ref{fig:expt4}.  Additionally, we estimate the theoretical value of the mean delay between two network endpoints using the M/M/c queue model for validation of experiment when using multi-core architecture. The latency values calculated are mean of five runs of each algorithm on a network of varying size. 

\noindent We observe latency of 0.45 for \texttt{network-size=50} - Figure~\ref{fig:expt4}(a). The theoretical estimate of 2-core OVS-DPDK architecture \texttt{SFC-Theoretical} is 0.28s. When using \texttt{SFC-NFP} the end to end latency is 0.23s. This shows a performance gain of 1.21x using our algorithm compared to the normal parallelized packet processing of VNF's. When the number of CPU cores is increased to 4, the latency for \texttt{network-size=50} - Figure~\ref{fig:expt4}(b) is 0.13s compared to 0.18s latency achieved using \texttt{SFC-Theoretical}. 

\noindent Similarly, \texttt{SFC-NFP} is able to obtain better latency value compared to serial or parallel methods when the number of CPU cores is increased to 8 - Figure~\ref{fig:expt4}(c). The performance gain is 1.21x-1.4x for \texttt{network-size=50} using \texttt{SFC-NFP}. This performance gain can be attributed to the fact that on flow rule compilation many flow table entries of IDS and NAT can be aggregated, which leads to faster packet delivery from source node to the destination. 
 
\noindent We increase the network size from 50 nodes to 250 nodes for each experiment i.e. Figure~\ref{fig:expt4}(a)-(c) to check the scalability of \texttt{SFC-NFP}. For 2 core OVS-DPDK network \texttt{network-size=100} the latency using \texttt{SFC-Theoretical} is 0.67s whereas latency achieved using \texttt{SFC-NFP} is 0.56s - Figure~\ref{fig:expt4}(a). When the \texttt{network-size=200} the latency using parallel packet processing \texttt{SFC-Theoretical} is 4.26s, whereas \texttt{SFC-NFP} has a latency of 2.72s which is $\sim$1.60x performance gain. Similarly, when \texttt{network-size=250}, the performance gain using our method is  $\sim$1.66x.

\noindent Our method of flow optimization scales well as we increase the number of network nodes with change in the number of CPU cores. For 4 core OVS-DPDK architecture - Figure~\ref{fig:expt4}(b), when \texttt{network-size=200}, the latency using \texttt{SFC-Theoretical} is 2.96s whereas the latency using \texttt{SFC-NFP} is 1.62s. Similarly, the latency for \texttt{network-size=250}, the latency for \texttt{SFC-Theoretical} is 6.1s whereas \texttt{SFC-NFP} achieves 4.05s latency, which is a  $\sim$1.60x performance gain.

\noindent The \texttt{network-size=200} for 8 core OVS-DPDK framework - Figure~\ref{fig:expt4}(c) shows slight increase in latency using \texttt{SFC-NFP} 1.82s compared to 1.62s that our algorithms achieved in 4 core architecture - Figure~\ref{fig:expt4}(b). This delay increase is due to overhead induced by multiple processors running simultaneously. For \texttt{network-size=250} our method again performs better compared to \texttt{SFC-Theoretical}. The latency for \texttt{SFC-Theoretical} is 4.1s whereas latency using \texttt{SFC-NFP} is 2.95s, a 1.35x performance gain. Overall, flow compression and aggregation help in reduction of flow table size and eventually latency in SDN enabled cloud network. The average performance gain is 1.40x and maximum gain is  $\sim$1.90x in some cases. 

\section{Related Work}
\noindent Parallelized packet processing has been introduced by ParaBox~\cite{parabox}. The authors exploit parallel packet processing opportunities across network functions. The output of packet processing after various operations is merged to ensure correct sequential processing of packets. Our approach introduces network function parallelism based SFC for security and performance enforcement on the cloud network. 
 
\noindent Prados \textit{et al}~\cite{prados2017analytical} have considered G/G/m based queuing model for checking mean response time for network functions for the 5G environment. The analysis is analytic in nature, comparing theoretical and simulated response time of Virtual Network Functions (VNF's). We validate our model using  M/M/c queue architecture in a cloud network, which was proposed as future work by authors Prados \textit{et al}~\cite{prados2017analytical}. Our framework shows a marked reduction in SFC latency compared to the value predicted by the M/M/c queue model for parallelized packet processing.

\noindent Distributed cloud security for SDN cloud has been discussed by Pisharody ~\textit{et al}~\cite{pisharody2016security},~\cite{pisharody2017brew}. The authors have identified conflicts in flow rules for SDN clouds. We have used an aggregation of flow rules as part of our SFC, however, there can sometimes be conflicting actions between flow rules when VNF's are compiled to SDN rules. We have not considered such cases in current work, but this can serve a future work for SFC in SDN enabled cloud network. Sun \textit{et al}~\cite{sun2017nfp} have utilized network function parallelism based on DPDK framework to improve the performance of NFV. Our work focuses on ordering security functions in network function parallelism for performance and security assessment.

%Analytical QoS assessment approaches $(AQA^2)$ has been proposed by Phan \textit{et al}~\cite{phanaqa} for maintaining SLA for QoS between servers and clients in a cloud network when network functions are deployed. The research work utilizes a single chain model with serial processing and mixed chain approach with load balancing in place for network functions. The authors have however not utilized parallelized packet processing as our work to enhance throughput and increase CPU utilization. Our M/M/c queue model considers parallelized VNF deployment.

\section{Conclusion and Future Work}
Optimal service function chaining is important for security enforcement and performance improvement in a cloud network. SDN allows compilation of various VNF's into the common format - OpenFlow. There is overlap in packet header space and action fields of various VNF's when SFC is provisioned in the cloud network. Our framework performs NFP in service function chaining by identifying the dependencies between various VNF's such as IDS and VPN. The service functions with independent action sets can be parallelized in order to reduce the performance overhead. Our framework \texttt{SFC-NFP} is able to achieve upto 1.90x performance gain on a large cloud network compared to parallelization on OVS-DPDK architecture in the absence of flow optimization. 
%The performance results based on Intel's evaluation of OVS with DPDK shows a gain in throughput~\cite{dpdk2}. This can be quite useful to deal with flooding attacks such as DDoS. Furthermore, a malicious attacker can use unordered VNF's to mount multi-hop attacks on the cloud network. We have not covered these scenarios in current work. We plan to evaluate SFC throughput for flooding attacks on our framework \texttt{SFC-NFP} and ability to handle SFC induced network attacks as a part of future work. 

\section*{Acknowledgment}
This research is based upon work supported by the NRL N00173-15-G017, NSF Grants 1642031, 1528099, and 1723440, and NSFC Grants 61628201 and 61571375.

\bibliographystyle{abbrv}
\bibliography{cns}

\end{document}